\documentclass[conference]{IEEEtran}

\usepackage{cite}
\usepackage{amsmath,amssymb}
\usepackage{booktabs}
\usepackage{graphicx}
\usepackage{url}
\usepackage{xcolor}

\newcommand{\system}{\textsc{EvidenceNet}}
\newcommand{\adfclaims}{\textsc{ADF-Claims}}
\ifdefined\SmartComMarkChanges
\newcommand{\revised}[1]{\textcolor{red}{#1}}
\else
\newcommand{\revised}[1]{#1}
\fi

\title{Can AI Agents Deliver Verifiable Network-Wide Outcomes Across Authority Boundaries?}

\author{
\IEEEauthorblockN{Tianzhu Zhang}
\IEEEauthorblockA{
Nokia Bell Labs\\
Massy, France\\
tianzhu.zhang@nokia-bell-labs.com
}
\and
\IEEEauthorblockN{Chih-Kai Huang}
\IEEEauthorblockA{
LTCI, Télécom Paris, Institut Polytechnique de Paris\\
Palaiseau, France\\
chih-kai.huang@telecom-paris.fr
}
\and
\IEEEauthorblockN{Meikang Qiu}
\IEEEauthorblockA{
Augusta University\\
Georgia, USA\\
qiumeikang@gmail.com
}
}

\begin{document}
\maketitle

\begin{abstract}
AI agents are increasingly involved in network automation, where they can initiate configuration changes through mediated operational interfaces and assess the resulting state. Nonetheless, operational networks usually span many devices and administrative domains. Realizing an operator's intent requires coordinating agents with distinct authority scopes that define the resources they can access, the operations they can invoke, and the network state they can observe. This division limits the blast radius of an erroneous action but fragments the evidence needed to assess the network-wide outcome. Successful execution of a configuration action proposed by one agent does not establish that remote devices responded as intended or that routing changes reached the required devices.
A valid observation may also become stale after a subsequent change. Before the coordinated operation can be declared complete, a trusted assurance layer must collect current observations from the required scopes and determine whether they collectively support the operator's intended network-wide outcome. 

To address the \emph{completion admission} problem, we present \system{}, a runtime assurance layer for deciding whether coordinated agent operations have achieved an operator's network intent. Its broker collects the post-change observations required by a completion contract, and its admission gate checks that the evidence comes from the required scopes, remains current, and satisfies the task rules. A verifier agent provides an additional assessment of the observation content. Experiments on live routing networks show that post-change state checks recognize successful outcomes that configuration-action records alone cannot establish. Controlled interventions further show that \system{} rejects completion when otherwise satisfactory observations have the wrong source, have been substituted, or are stale. The collected evidence can also guide recovery from configuration faults. These results support an emerging vision of autonomous networking in which agents drive planning, configuration, interpretation, and repair, while a trusted runtime layer controls the evidence and machine-checkable conditions used to admit completion.
\end{abstract}


\section{Introduction}

Over the past several decades, network operations have evolved from manual, device-by-device configuration to scriptable management and controller-based automation. This evolution shifted operators from issuing device-level commands to specifying desired network outcomes, a shift formalized by intent-based networking (IBN)~\cite{rfc9315}. Once an automation system translates an intent into network actions, it must also assess whether the resulting state fulfills that intent. Service assurance closes this loop by monitoring the network and initiating corrective action when the delivered service deviates from the intent~\cite{rfc9417}. Conventional IBN implementations generally rely on predefined workflows, deterministic control logic, and explicitly engineered checks. LLM-powered AI agents provide a more flexible approach. They can interpret natural-language intent, compose multistep operations, invoke heterogeneous operational tools, and revise their plans in response to observed network state~\cite{mani2023llmnet,zhou2025meshagent,asadli2026repair,zhao2026nma}. As these agents move from advisory support to direct network actuation, their operational authority must be regulated and, whenever applicable, constrained~\cite{zhang2026network}.

Operational networks span many devices and may cross functional or administrative boundaries. Granting one agent unrestricted authority over such a network would weaken administrative separation and enlarge the impact of an erroneous action. Network-wide tasks are thus coordinated among agents with distinct \emph{authority scopes}, each specifying the resources an agent can access, the operations it can invoke, and the network state it can directly observe~\cite{zhao2026nma,wang2026agentspec,ji2026seagent}. These scopes limit the blast radius of erroneous actions, but also divide direct observation. A configuration accepted within one scope does not establish that a peer's behavior is consistent, an expected route has successfully propagated, or a policy has taken effect at the required network locations.

\begin{figure}[!tb]
\centering
\includegraphics[width=0.5\textwidth]{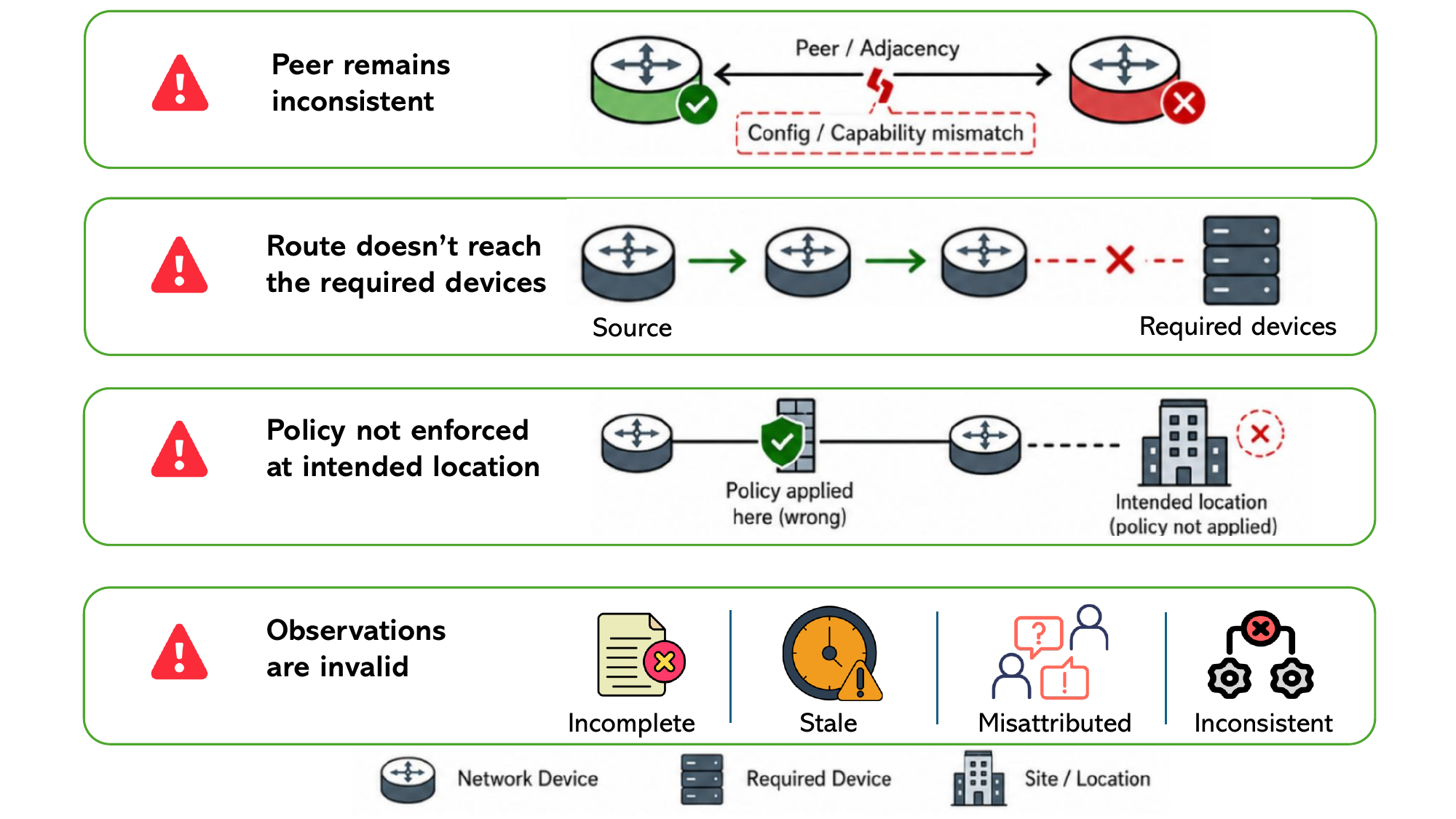}
\caption{Local agent success does not establish a network-wide outcome.}
\label{fig:problem}
\end{figure}

This observation gap becomes consequential when the automation system declares a coordinated operation complete. Such a declaration may activate a service, migrate traffic, remove a fallback path, or authorize a dependent configuration change. Therefore, the network operator must specify the observations required to establish the intended outcome, while the automation system must determine whether those observations are complete, correctly sourced, and up to date~\cite{zhou2011secure}. Fig.~\ref{fig:problem} illustrates this assurance problem.

Existing assurance architectures commonly rely on a trusted component with direct access to the relevant network state~\cite{rfc9417,fogel2015batfish,khurshid2013veriflow}. When authority and visibility are divided across agents, however, no individual agent can establish the overall outcome from its local observations alone. The system must instead combine evidence collected from the required scopes and decide whether it collectively supports the operator's intent. We call this decision \emph{completion admission}. An overly permissive decision may advance the workflow without sufficient support, whereas an overly conservative one may reject a correct outcome and block autonomous progress.

We propose \system{}, a trusted runtime assurance layer for completion admission among scope-constrained network agents. Agents plan operations, propose configurations, interpret observations, and suggest repairs, but they cannot declare their own success. \system{} collects the observations required by a completion contract, binds them to their source and network epoch, and admits completion only when deterministic checks and the verifier assessment both succeed. It complements existing planners and network property checkers by controlling whether the available evidence is sufficient for the coordinated workflow to continue. The main contributions of this paper are as follows:
\begin{itemize}
    \item We formulate \emph{completion admission} for network operations spanning multiple authority scopes and define a trusted completion contract that specifies the observations required to establish each required network property
    \item We design and implement \system{}, which mediates scoped agent actions, collects provenance- and freshness-bearing observations, and enforces their coverage and validity before completion may be admitted.
    \item We evaluate \system{} through live-network experiments and controlled evidence interventions, separating the effects of post-change evidence acquisition, deterministic evidence controls, verifier judgment, and repair.
\end{itemize}


\section{Related Work}

Traditional network assurance and verification address complementary parts of operational correctness. Assurance mechanisms monitor whether a delivered service continues to satisfy the operator's intent and may trigger corrective action when it does not~\cite{rfc9315,rfc9417}. Network verification, by contrast, checks whether configurations or forwarding state satisfy specified invariants. In particular, Batfish analyzes forwarding behavior from network-wide configurations~\cite{fogel2015batfish}, while VeriFlow verifies invariants as controller rules change~\cite{khurshid2013veriflow}. \system{} complements both lines of work. Instead of introducing new network properties or verification algorithms, it ensures that the observations supporting those properties originate from the required authority scopes, stay up to date, and collectively support the intended network-wide outcome.

Recent work explores LLMs across several stages of network management, including executable analysis~\cite{mani2023llmnet}, configuration synthesis with verifier feedback~\cite{mondal2023router}, invariant-guided workflows~\cite{zhou2025meshagent}, configuration repair~\cite{asadli2026repair}, and contract-governed intent translation~\cite{bimo2026caif}. These works improve task-level correctness, but do not directly address completion assurance when network control and observation are divided among agents with distinct authority scopes. CAIF~\cite{bimo2026caif} is architecturally closest to \system{} as it separates probabilistic reasoning from network actuation through a machine-readable contract. However, CAIF validates intents and policies before actuation, whereas \system{} determines whether observations collected across authority scopes provide sufficient support to accept a coordinated operation as complete.

Some works govern what agents may do or make their executions traceable.
AgentSpec enforces runtime policies for tool-using agents~\cite{wang2026agentspec}, while SEAgent applies mandatory access control to agent privileges~\cite{ji2026seagent}. Execution-provenance research studies how agent actions and tool invocations produce their outputs~\cite{wang2026provenance}. They do not address whether observations collected after a coordinated change come from the required authority scopes, remain valid for the current network, and collectively provide sufficient support to declare completion.


\begin{figure*}[!tb]
\centering
\includegraphics[width=0.95\textwidth]{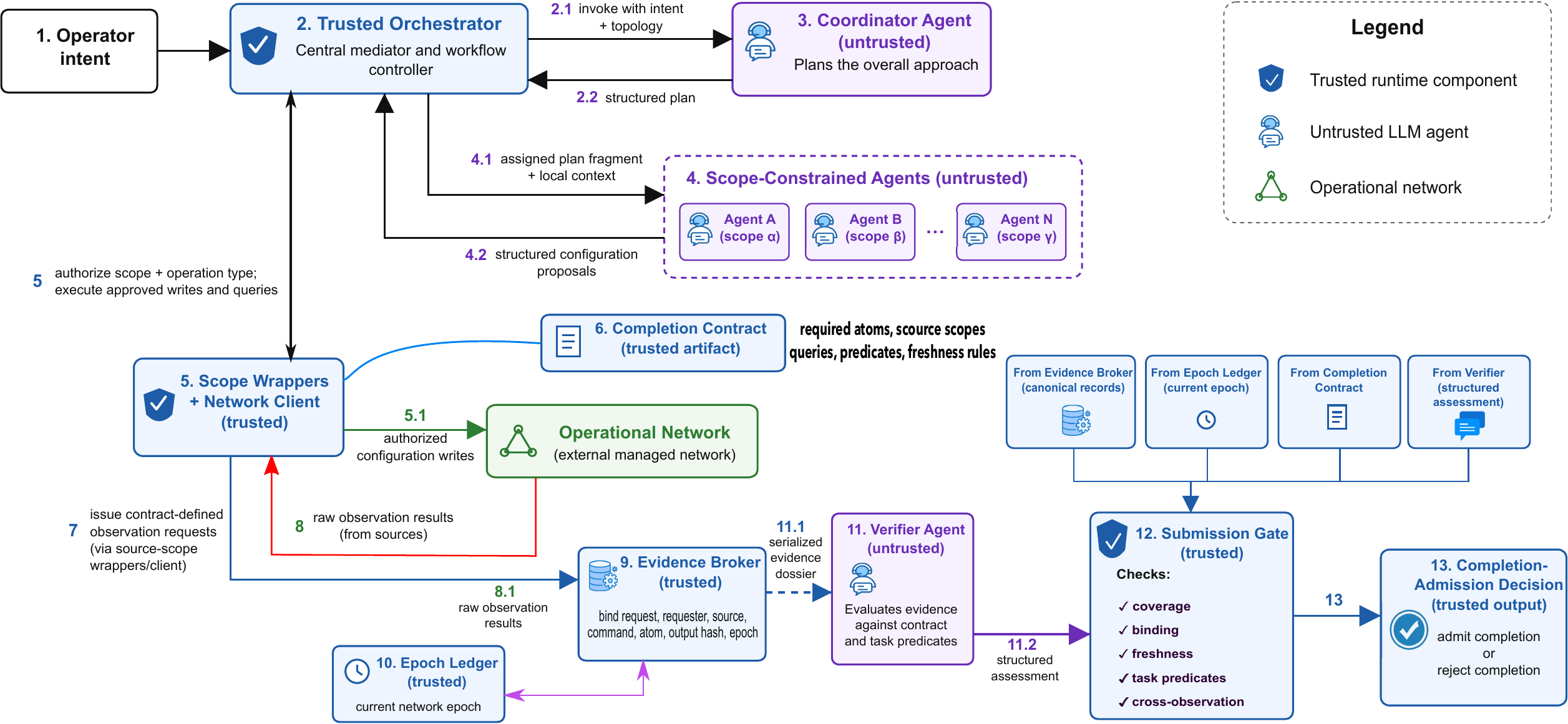}
\caption{\system{} architecture and its workflow for trusted evidence collection and completion admission.}
\label{fig:architecture}
\end{figure*}

\section{EvidenceNet Design}\label{sec:design}

\system{} separates agent planning and interpretation from trusted action mediation, evidence collection, and completion admission.



\subsection{Completion contract}

Let $\mathcal{P}$ denote the network properties required by an operator's intent. An observation atom is the smallest required unit of evidence that contributes to checking one such property. Each atom specifies a source location, an exact query, a deterministic parser with
success condition, and a freshness rule. It may also identify a requester scope and relationships with other atoms. Reciprocal adjacency, for example, requires two atoms because each endpoint must observe the other. Each property $p$ maps to a set of atoms $\mathcal{O}(p)$, and $\mathcal{O}=\bigcup_{p\in\mathcal{P}}\mathcal{O}(p)$. This mapping forms the completion contract.

For each atom $o\in\mathcal{O}$, the orchestrator registers an observation request and instructs the wrapper for the atom's source scope to execute the declared query. The \emph{evidence broker} converts the returned result into an evidence record $e_o$. The record
contains its issued identity, request identifier, requester and collector identities, source router, command, execution status, complete returned output and its hash, atom identifier, collection time, and network epoch. The broker derives these fields from the mediated operation; agent-generated text cannot create or alter a broker-issued record. The records for a completion decision form an \emph{evidence dossier} $D$.

The submission gate follows a fail-closed policy. Admission requires coverage of every declared atom, successful deterministic checks, satisfaction of declared relationships between
observations, explicit per-atom support from the verifier, as well as a positive global recommendation:
\begin{equation}
\begin{aligned}
A(D)={}&C(D,\mathcal{O})\land X(D,\mathcal{O})\land V_g(D)\\
&{}\land\bigwedge_{o\in\mathcal{O}}
\left[B(e_o,o)\land F(e_o)\land Q(e_o,o)
\land S_V(D,o)\right].
\end{aligned}
\label{eq:gate}
\end{equation}
$C$ requires a usable broker-issued record for every declared atom and requires the verifier response to account for each atom exactly once
as supported or unsupported. Requests containing undeclared atom identifiers are rejected. 
$B$ verifies that a record matches its issued identity, request, requester, source, command, atom, and output hash. $F$ verifies that the record remains current. $Q$ applies the atom's parser and success condition, called its \emph{task predicate}. $X$ applies any declared
deterministic relationships between records, such as equality of a community observed at two routers. These terms are deterministic. The verifier reports per-atom support through $S_V$ and returns the global recommendation $V_g$. A positive verifier recommendation is
necessary but cannot override a deterministic failure.
The internal admission decision $A$ remains separate from the external evaluator outcome $Y$ for the encoded task properties. For $A,Y\in\{0,1\}$, the pair $(A,Y)$ distinguishes supported success $(1,1)$, false admission $(1,0)$, false rejection $(0,1)$, and justified rejection $(0,0)$.

\subsection{Orchestration, scoped execution, and evidence collection}

The architecture and control flow of \system{} are illustrated in Fig.~\ref{fig:architecture}. 
\system{} comprises a set of deterministic, trusted software components to enforce its fixed completion contract, including the orchestrator, scope wrappers, evidence broker, epoch ledger, network client, and submission gate. It also encompasses a collection of network agents with different roles, including the coordinator, verifier, repairer, and multiple scope-constrained agents. However, these agents are not trusted components. They can produce plans, proposals, and assessments, but cannot modify the contract or bypass a failed admission check. We assume that the orchestrator, network client, and managed routers are not compromised. The prototype does not detect configuration changes that bypass the mediated interfaces.

The coordinator agent receives the operator's intent and relevant network context, then decomposes the operation into plans aligned with the participating authority scopes. For each scope, the orchestrator invokes a scope-constrained agent with the local context and assigned portion of the plan. The agent returns structured configuration proposals for that scope.

The orchestrator then sends each proposal to a scope wrapper configured with the agent's authority scope. The wrapper checks the target router and operation type before forwarding an authorized operation to the network client. Thus, an agent assigned to Scope 1 may propose actions for that scope, but its wrapper rejects operations targeting other scopes. The wrapper also rejects a mutating command presented as a read operation. 

After the operation is actuated, the orchestrator iterates over the observation atoms in the completion contract. The requester identifies the scope for which an observation is required, whereas the source identifies the only scope authorized to collect it.

For each atom, the orchestrator registers an evidence request under the declared requester identity. The evidence broker verifies that the
request matches the contract, and the orchestrator then asks the source scope's wrapper to fulfill it. The source wrapper executes the declared query through the network client. The broker converts the returned result into the evidence record defined above.
The broker maintains router-specific epochs and a global network epoch. Every successful mediated configuration command advances the global epoch. The evaluated prototype adopts a conservative global freshness policy, under which any observation collected at an earlier epoch is stale. Such observations must be recollected before completion can be admitted.

\subsection{Verifier assessment and one-round repair}

After evidence collection, the orchestrator constructs a serialized dossier from the completion contract and the broker's records. The
verifier agent receives the operator's intent, the dossier, and its freshness metadata. It identifies supported and unsupported atoms and returns a structured assessment containing per-atom decisions and a global recommendation.

Upon receiving the verifier assessment, the orchestrator invokes the submission gate defined in Eq.~\eqref{eq:gate} over the broker's canonical records and the current network epoch. The submission wrapper invokes the same gate once more immediately before submission.


If admission fails, \system{} may attempt at most one repair round. We allow only one round to avoid repeated, potentially harmful changes. If admission still fails, the workflow stops. A repair coordinator agent receives the failed atoms, relevant evidence, candidate authority scopes, and the original task. It proposes a coordinated repair plan. The orchestrator assigns the corresponding steps to the affected scope-constrained agents, which return structured repair actions. Every action passes through the same router-specific wrapper and authorization checks as the initial execution.

After the repair actions stabilize, the orchestrator collects a new observation for each atom in the completion contract. Any successful repair command advances the global epoch, making the earlier dossier stale under the prototype's freshness policy. The orchestrator then constructs a new dossier, invokes the verifier, and calls the submission gate as before. 


\section{Evaluation}

We organize the evaluation around four research questions.
\begin{itemize}
\item \textbf{Q1:} Does post-change evidence enable successful outcomes to be admitted while failures remain rejected?
\item \textbf{Q2:} Do source-binding and epoch-based freshness checks reject wrong-source, substituted, and stale evidence?
\item \textbf{Q3:} Can the LLM-powered verifier agent reject defects that are visible in the observation content but not encoded by the deterministic predicate checks?
\item \textbf{Q4:} Can one-round repair successfully restore the intended network outcome, and how precisely does it target the affected network elements?
\end{itemize}

\begin{table}[b]
\caption{Live-network outcomes and completion admissions. Entries are numbers of traces.}
\label{tab:action-results}
\centering
\scriptsize
\setlength{\tabcolsep}{3pt}
\begin{tabular}{lrrrr}
\toprule
\textbf{Task family} &
\textbf{Traces} &
\shortstack{\textbf{Evaluator}\\\textbf{success}} &
\shortstack{\textbf{\adfclaims{}}\\\textbf{admit}} &
\shortstack{\textbf{\system{}}\\\textbf{admit}} \\
\midrule
Reciprocal OSPF & 15 & 10 & 0 & 10 \\
BGP propagation & 15 & 10 & 0 & 10 \\
BGP filtering & 15 & 12 & 0 & 12 \\
\midrule
All traces & 45 & 32 & 0 & 32 \\
\bottomrule
\end{tabular}
\vspace{-5mm}
\end{table}

We implement a prototype of \system{} on NetAgentBench~\cite{twabi2026netagentbench}, which provides the basic runtime environment and a separate evaluator to score the resulting live network. The evaluator is unavailable to the agents and lies outside the completion-admission path. We use it as the external evaluator. All live-network experiments run in Containerlab using FRRouting, with the router image pinned to \texttt{frrouting/frr:v8.4.1}. The runtime and library implementation add circa 10,000 lines of code to NetAgentBench. The prototype instantiates each authority scope as one router and associates it with one scope-constrained agent.


We evaluate \system{} on six custom tasks implemented in the NetAgentBench task format. They cover reciprocal OSPF adjacency and
loopback reachability, multi-area OSPF, BGP route propagation, selective route filtering, BGP community and path policies, and reachability policies. Each task specifies an agent-visible objective and topology, a completion contract, and model-hidden checks for evaluating the final network state. Unless otherwise specified, all agents use \texttt{gpt-5.4-mini} as the backend LLM.



\subsection*{Q1: evidence acquisition}
To analyze the impact of post-change evidence, we generated 45 fulfillment traces, 15 from each of the reciprocal OSPF, BGP propagation, and BGP filtering tasks. Each trace records the ordered router commands executed during an agent run and their immediate results, but contains no post-change observations or evaluator verdict. We replayed each command sequence on two fresh instances of the same topology. The Action-Derived Fulfillment Claims (\adfclaims{}) baseline assessed completion from the execution records alone, whereas \system{} collected the observations required by the completion contract before making its admission decision.
After each replay, the external evaluator checked the resulting live network separately. Across the 45 traces, 32 produced a network state that satisfied every requirement of the corresponding task, while 13 failed at least one requirement. 

Table~\ref{tab:action-results} shows that \system{} agreed with the evaluator in all 45 cases. \adfclaims{} correctly rejected the 13 failures but also rejected all 32 successful outcomes. Acceptance alone thus does not establish the intended network state.

\subsection*{Q2: Binding and freshness}

We reuse one successful Q1 trace from each of the three tasks. We replay each trace three times under five conditions: clean evidence, evidence from the wrong router, a substituted record, stale evidence retained after a network fault, and current evidence collected after that fault. The wrong-source, substituted-record, and stale conditions test whether apparently satisfactory observations remain admissible when their provenance or freshness is invalid.

We compare \system{} with two alternatives. The \emph{content-only baseline} judges whether the visible observation content satisfies the task, without checking its brokered origin or freshness. The \emph{verifier-free variant} retains \system{}'s deterministic admission checks but omits the verifier. These comparisons separate the contribution of broker-enforced evidence controls from that of the verifier.
According to the results in Table~\ref{tab:evidence-results}, all three methods accept the 9 clean-evidence cases and reject the 9 current-fault cases. For the 27 wrong-source, substituted, or stale cases, the content-only baseline admits every dossier, whereas the deterministic gate and \system{} reject all of them. Their identical decisions show that the verifier agent adds no further separation in this experiment.


\begin{table}[!tb]
\caption{Admission decisions under controlled evidence conditions. Entries report dossiers admitted out of nine.}
\label{tab:evidence-results}
\centering
\scriptsize
\setlength{\tabcolsep}{1.8pt}
\begin{tabular}{lccccc}
\toprule
\textbf{Condition} &
\shortstack{\textbf{Network}\\\textbf{evaluator}} &
\shortstack{\textbf{Evidence}\\\textbf{state}} &
\shortstack{\textbf{Content-only}\\\textbf{baseline}} &
\shortstack{\textbf{Verifier-free}\\\textbf{variant}} &
\textbf{\system{}} \\
\midrule
Clean & pass & valid & 9/9 & 9/9 & 9/9 \\
Wrong source & pass & wrong source & 9/9 & 0/9 & 0/9 \\
Substituted record & pass & substituted & 9/9 & 0/9 & 0/9 \\
Stale after change & fail & stale & 9/9 & 0/9 & 0/9 \\
Fresh after change & fail & current & 0/9 & 0/9 & 0/9 \\
\bottomrule
\end{tabular}
\end{table}

\subsection*{Q3: verifier assessment}
We begin with six clean dossiers derived from six successful task contexts. For each dossier, we retain the clean version and create four altered variants, yielding six clean and 24 altered dossiers. Each altered version contains one visible defect: an outdated snapshot, output attributed to an undeclared command, an observation contradicted by a later observation, or a monitoring summary presented as raw device output. The altered dossiers remain correctly formatted and pass the deterministic gate because these four defects are not encoded in its existing rules.

The two verifiers (backed by GPT-5.4 mini and GPT-5.3 Codex Spark) assess every dossier three times. We hide the condition labels, use neutral identifiers, randomize the order, and isolate every call. The repetitions measure whether an agent makes the same decision consistently. Before replay, we fixed a narrow rule-based checker for the four planted defect classes. We hid it from the verifiers and did not include it in the operational \system{} gate.

As shown in Table~\ref{tab:semantic-results}, both verifiers accept all 6 clean dossiers, but they also accept all 24 altered dossiers. The rule-based checker accepts all 6 clean dossiers and rejects all 24 altered dossiers.
This result suggests that deterministic rules should enforce known, precisely defined defects rather than leaving them to verifier judgment. However, the rule-based checker recognizes only the four defects studied here and is not a general solution for interpreting arbitrary network evidence.

\begin{table}[!tb]
\caption{Completion decisions in the blinded evidence-defect study.}
\label{tab:semantic-results}
\centering
\scriptsize
\setlength{\tabcolsep}{2pt}
\begin{tabular}{p{0.33\columnwidth}p{0.14\columnwidth}p{0.18\columnwidth}p{0.22\columnwidth}}
\toprule
\textbf{Decision path} &
\textbf{Clean admit} &
\textbf{Altered dossiers rejected} &
\textbf{Repeat stability} \\
\midrule
Full gate + GPT-5.4 mini & 6/6 & 0/24 & 30/30 stable \\
Full gate + GPT-5.3 Spark & 6/6 & 0/24 & 30/30 stable \\
Rule-based checker & 6/6 & 24/24 & deterministic \\
\bottomrule
\end{tabular}
\end{table}

\begin{table}[b]
\caption{External outcome with and without one repair round.}
\label{tab:repair-results}
\centering
\small
\begin{tabular}{lrr}
\toprule
\textbf{Fault} & \textbf{No repair} & \textbf{One round} \\
\midrule
BGP neighbor setting & 0/3 & 3/3 \\
BGP filter attachment & 0/3 & 3/3 \\
OSPF network type & 0/3 & 3/3 \\
\midrule
All cases & 0/9 & 9/9 \\
\bottomrule
\end{tabular}
\end{table}

\subsection*{Q4: one-round repair}
We inject three faults, each affecting one router: an incorrect BGP neighbor AS, an incorrectly attached route filter, and an incorrect OSPF network type. We test each fault three times with repair disabled and three times with one repair round allowed, for 18 runs in total.
With repair disabled, all nine runs failed both the external evaluator and completion admission. With one repair round, all nine runs passed both checks.
We also find that repairs were successful but broader than necessary, and every repair issued configuration commands to all three routers.

\section{Limitations}
The current \system{} prototype demonstrates the architecture, but it still bears several limitations.  

\subsection{Trust and enforcement boundary}

The prototype assumes that its orchestrator, wrappers, broker, network client, gate, and observation sources are not compromised. Managed routers may be misconfigured, but they are assumed not to falsify observations. In-process wrappers enforce authority scopes rather than operating-system isolation or cryptographic capabilities. Broker records establish request, source, command, and content-hash consistency within the trusted runtime, but do not authenticate a router's actual state or detect changes that bypass the wrappers. The one-router-per-scope design also does not cover more complex administrative or functional scopes.

\revised{The controlled substitutions therefore test misuse of records issued by an honest broker, not observation forgery by a compromised source. A compromised router could report a plausible but false state, and a compromised broker or gate could bypass admission. Stronger deployment would require isolated enforcement, authenticated collectors, protected logs, and independent observations for critical properties. Signatures can authenticate an issuer and detect later modification, but cannot establish that the issuer reported the network state truthfully. These protections are not evaluated here.}

\subsection{Freshness and contract coverage}

The global epoch invalidates all evidence after every successful configuration command. This is conservative because unrelated evidence may be recollected, yet changes outside the wrappers may remain invisible. Observations are also collected sequentially, so they do not form an atomic network snapshot, and the fixed stabilization wait does not prove convergence.

\revised{A finer-grained extension would associate each observation with the resources that can invalidate it and their versions. For example, a routing-policy change may invalidate route and reachability observations at downstream routers while leaving an unrelated adjacency observation usable. The gate would recollect affected observations and recheck their versions before admission. This requires complete tracking of indirect effects, including route propagation; when dependencies are uncertain, global invalidation remains the conservative fallback. Our experiments do not measure the cost or safety of this extension.
}
Completion admission covers only the properties represented in the completion contract. An omitted requirement or weak predicate may therefore allow an incomplete outcome to be admitted. The verifier cannot establish a property for which it collected no evidence.
\revised{The experiments establish no positive benefit from LLM interpretation of unstructured observations. The verifier is an additional assessment whose benefit remains unproven; it cannot override a deterministic rejection. Testing that benefit would require independently labeled, unfamiliar observations with genuinely unresolved interpretation, rather than defects with known exact checks.}

\subsection{Repair and evaluation scope}

The repair study covers one round and three single-router faults. It does not provide minimal changes, staged deployment, rollback, or guarantees under simultaneous faults.

\revised{Recovery success and targeting precision are distinct. Counting routers receiving configuration commands, the faulty router was always included, but only one of the three targeted routers was known to require repair. Some commands may merely reapply existing configuration, so this does not measure the number of changed settings. A proposed improvement is to diagnose the failed observations, authorize an explicit target and operation set, inspect the configuration difference, and reverify affected and previously satisfied properties. Minimality would require a separate objective and evaluation; the current one-round bound limits repetition, not the change set.}

The evaluation uses Containerlab, one FRRouting release, small topologies, six tasks, and two verifier backends. Repeated verifier calls over the same dossiers measure decision consistency rather than generalization to independent tasks. The Q2 intervention repetitions also share a small number of underlying network contexts. The \adfclaims{} baseline receives no post-change observations, so the comparison isolates the value of evidence acquisition. The controlled interventions demonstrate enforcement, and the results do not establish behavior across vendors, substrate networks, larger deployments, adversarial settings, or concurrent changes.
\revised{The general abstraction permits a scope to contain several devices, but the singleton-scope prototype does not validate this mapping. Larger networks would increase both cross-scope observations and their dependencies. Generalization should therefore be tested by independently varying devices per scope, participating scopes, routing diversity, and concurrent changes, and measuring admission errors, collection cost, and completion latency. The current results establish the mechanism on the tested tasks, not scalability to such environments.}


\section{Conclusion}
Autonomous network operation requires more than successful actuation. Before a coordinated workflow can advance, the resulting network state must be established from current observations collected at the relevant authority scopes. This paper introduced \system{}, a runtime assurance architecture that separates agent-driven configuration from completion admission.

Our prototype evaluation shows that post-change evidence can admit successful outcomes that configuration-action records alone cannot establish. Source binding and freshness checks reject observations that appear satisfactory but originate from the wrong location, have been substituted, or no longer describe the current network state. One-round repair restored every evaluated fault, although the resulting changes were broader than necessary. Furthermore, two verifier agents backed by different LLMs failed to detect the injected content defects that explicit rules recognized, demonstrating that agent agreement is not itself a reliable assurance boundary.
As network agents gain greater operational autonomy, preserving this separation between intelligent action and trusted assurance is essential to verifiable outcomes.


\bibliographystyle{IEEEtran}
\bibliography{evidencenet_smartcom26}

\end{document}